\documentclass[twocolumn]{aastex62}

\usepackage{float}

\graphicspath{{./}{myfigures/}}

\submitjournal{AAS Journals}

\shorttitle{OGLE-2018-BLG-0677}
\shortauthors{Herrera-Mart\'in A. et al.}

\begin{document}

\title{OGLE-2018-BLG-0677Lb: A SUPER EARTH NEAR THE GALACTIC BULGE }

%\correspondingauthor{Antonio}
%\email{greg.schwarz@aas.org, gus.muench@aas.org}

\author[0000-0002-3654-4662]{Antonio Herrera-Mart\'in}
\affil{School of Physical and Chemical Sciences, University of Canterbury, Private Bag 4800, Christchurch, New Zealand}

\author[0000-0003-3316-4012]{M. D. Albrow}
\affiliation{School of Physical and Chemical Sciences, University of Canterbury, Private Bag 4800, Christchurch, New Zealand}

\author{A. Udalski }
\affiliation{Astronomical Observatory, University of warsaw, Al. Ujazdowskie 4, 00-478 Warszawa, Poland}

\author{A. Gould}
\affiliation{Department of Astronomy, Ohio State University, 140 W.18th Ave., Columbus, OH 43210, USA}
\affiliation{Korea Astronomy and Space Science Institute, Daejon 34055, Korea}
\affiliation{Max-Planck-Institute for Astronomy, K\"{o}nigstuhl 17, 69117 Heidelberg, Germany}

\author{Y.-H. Ryu}
\affiliation{Korea Astronomy and Space Science Institute, Daejon 34055, Korea}

\author{J. C. Yee}
\affiliation{Harvard-Smithsonian Center for Astrophysics, 60 Garden St., Cambridge, MA 02138, USA}

\nocollaboration

%
%  KMT
%
\author{S.-J. Chung}
\affiliation{Korea Astronomy and Space Science Institute, Daejon 34055, Korea}
\affiliation{Korea University of Science and Technology, 217 Gajeong-ro, Yuseong-gu, Daejeon, 34113, Korea}

\author{C. Han}
\affiliation{Department of Physics, Chungbuk National University, Cheongju 28644, Korea}

\author{K.-H. Hwang}
\affiliation{Korea Astronomy and Space Science Institute, Daejon 34055, Korea}

\author{Y. K. Jung}
\affiliation{Harvard-Smithsonian Center for Astrophysics, 60 Garden St., Cambridge, MA 02138, USA}

\author{C.-U. Lee}
\affiliation{Korea Astronomy and Space Science Institute, Daejon 34055, Korea}
\affiliation{Korea University of Science and Technology, 217 Gajeong-ro, Yuseong-gu, Daejeon, 34113, Korea}

\author{I.-G. Shin}
\affiliation{Harvard-Smithsonian Center for Astrophysics, 60 Garden St., Cambridge, MA 02138, USA}

\author{Y. Shvartzvald}
\affiliation{Department of Particle Physics and Astrophysics, Weizmann Institute of Science, Rehovot 76100, Israel}

\author[0000-0001-6000-3463]{W. Zang} 
\affiliation{Department of Astronomy and Tsinghua Centre for Astrophysics, Tsinghua University, Beijing 100084, China}

\author{S.-M. Cha}
\affiliation{Korea Astronomy and Space Science Institute, Daejon 34055, Korea}
\affiliation{School of Space Research, Kyung Hee University,Yongin, Kyeonggi 17104, Korea}

\author{D.-J. Kim}
\affiliation{Korea Astronomy and Space Science Institute, Daejon 34055, Korea}

\author{H.-W. Kim}
\affiliation{Korea Astronomy and Space Science Institute, Daejon 34055, Korea}

\author{S.-L. Kim}
\affiliation{Korea Astronomy and Space Science Institute, Daejon 34055, Korea}
\affiliation{Korea University of Science and Technology, 217 Gajeong-ro, Yuseong-gu, Daejeon, 34113, Korea}

\author{D.-J. Lee}
\affiliation{Korea Astronomy and Space Science Institute, Daejon 34055, Korea}
\affiliation{Korea University of Science and Technology, 217 Gajeong-ro, Yuseong-gu, Daejeon, 34113, Korea}

\author{Y. Lee}
\affiliation{Korea Astronomy and Space Science Institute, Daejon 34055, Korea}
\affiliation{School of Space Research, Kyung Hee University,Yongin, Kyeonggi 17104, Korea}

\author{B.-G. Park}
\affiliation{Korea Astronomy and Space Science Institute, Daejon 34055, Korea}
\affiliation{Korea University of Science and Technology, 217 Gajeong-ro, Yuseong-gu, Daejeon, 34113, Korea}

\author{R. W. Pogge}
\affiliation{Department of Astronomy, Ohio State University, 140 W.18th Ave., Columbus, OH 43210, USA}

\collaboration{(KMTNet Collaboration)}

%
% OGLE
%

\author{M. K. Szyma\'{n}ski}
\affiliation{Astronomical Observatory, University of warsaw, Al. Ujazdowskie 4, 00-478 Warszawa, Poland}

\author{P. Mr\'{o}z}
\affiliation{Division of Physics, Mathematics, and Astronomy, California Institute of Technology, Pasadena, CA 91125, USA}

\author{J. Skowron}
\affiliation{Astronomical Observatory, University of warsaw, Al. Ujazdowskie 4, 00-478 Warszawa, Poland}

\author{R. Poleski}
\affiliation{Astronomical Observatory, University of warsaw, Al. Ujazdowskie 4, 00-478 Warszawa, Poland}
\affiliation{Department of Astronomy, Ohio State University, 140 W.18th Ave., Columbus, OH 43210, USA}

\author{I. Soszy\'{n}ski}
\affiliation{Astronomical Observatory, University of warsaw, Al. Ujazdowskie 4, 00-478 Warszawa, Poland}

\author{S. Koz{\l}owski}
\affiliation{Astronomical Observatory, University of warsaw, Al. Ujazdowskie 4, 00-478 Warszawa, Poland}

\author{P. Pietrukowicz}
\affiliation{Astronomical Observatory, University of warsaw, Al. Ujazdowskie 4, 00-478 Warszawa, Poland}

\author{K. Ulaczyk}
\affiliation{Astronomical Observatory, University of warsaw, Al. Ujazdowskie 4, 00-478 Warszawa, Poland}

\author{K. Rybicki}
\affiliation{Astronomical Observatory, University of warsaw, Al. Ujazdowskie 4, 00-478 Warszawa, Poland}

\author{P. Iwanek}
\affiliation{Astronomical Observatory, University of warsaw, Al. Ujazdowskie 4, 00-478 Warszawa, Poland}

\author{M. Wrona}
\affiliation{Astronomical Observatory, University of warsaw, Al. Ujazdowskie 4, 00-478 Warszawa, Poland}

\collaboration{(OGLE Collaboration)}

%% Mark off the abstract in the ``abstract'' environment. 
\begin{abstract}

We report the analysis of the microlensing event OGLE-2018-BLG-0677. A  
small feature in the light curve of the event leads 
to the discovery that the lens is a star-planet system. Although
there are two degenerate solutions that could not be distinguished for this event, both
lead to a similar planet-host mass ratio.
We perform a Bayesian analysis based on a Galactic
model to obtain
the properties of the system and find that the planet corresponds to
a super-Earth/sub-Neptune with a mass $M_{\mathrm{planet}} = {3.96}^{+5.88}_{-2.66}\mathrm{M_\oplus}$.
The host star has a mass 
$ M_{\mathrm{host}} = {0.12}^{+0.14}_{-0.08}\mathrm{M_\odot}$.
The projected separation for the inner and outer solutions 
are ${0.63}^{+0.20}_{-0.17}$~AU and ${0.72}^{+0.23}_{-0.19}$~AU respectively. At
$\Delta\chi^2=\chi^2({\rm 1L1S})-\chi^2({\rm 2L1S})=46$, this is by far the lowest $\Delta\chi^2$ for any securely-detected microlensing planet to date, a feature that is closely connected to the fact that it is {detected primarily via a ``dip'' rather than a ``bump''.}

% \footnote{Note that manuscripts 
% submitted to the new Research Notes of the American Astronomical Society 
% (RNAAS) do \textbf{not} have abstracts.}.  If you exceed this length the
% Editorial office will ask you to shorten it.

\end{abstract}

%% Keywords should appear after the \end{abstract} command. 
%% See the online documentation for the full list of available subject
%% keywords and the rules for their use.

\section{INTRODUCTION} \label{sec:intro}
In the study of astronomical bodies, the search for extra-solar planets is of particular
interest as their characterization not only allows us to infer the similarities or 
differences in the mechanisms of their formation, but it also helps us to better
understand our own solar system~\citep{ollivier2008planetary}. There is a wide range of methods for planet 
discovery such as {radial velocity (RV)}, transit photometry, microlensing, direct 
imagining, etc. This has been a huge leap since the first confirmed discovery of
an extra-solar planet more than 20 years ago~\citep{Cochran1991Constraints114762}; to date
there are four thousand extra-solar planets ($\simeq$ 4104 as of 
December 2019), {with the majority ($\geq 3000$) having been} found by the transit method.

Gravitational microlensing is a particular type of gravitational lensing for which both
source and lens are stellar mass {objects and the angular} size of the magnified 
images cannot be resolved~\citep[etc]{vietriostriker1983, Paczynski1986GravitationalHalo, schneider2006gravitational,Tsapras2018MicrolensingExoplanets}. Instead, we 
study the difference in brightness of the source produced by the gravitational 
interaction of the lens when it crosses through the line of sight~\citep[etc]{Einstein506, Paczynski1986GravitationalHalo, schneider2013gravitational,mollerach2002gravitational,Mao2012AstrophysicalMicrolensing}. Although, it was
theorized since the formulation of gravitational lensing, the very small likelihood of the
necessary alignment discouraged its observation as having low probability. The
proposition
had a resurgence in interest following the work { of~\cite{Paczynski1986GravitationalHalo}, which} led
to the start of microlensing observations with the first detection in 
1993~\citep{Alcock1993PossibleCloud, Udalski:1993zz}.

\cite{Mao1991GravitationalSystems} pointed out how the formalism for binary
lenses and the feasibility of their observation could be used for detecting either binary systems or a 
planetary system. This was developed by several authors in 
the following years~\citep[etc]{Gouldloeb:1992,Griest1992EffectMicrolensing, Gould2000AMicrolensing, Albrow2000Detection97BLG41, Bennett2002GravitationalHoles,An2002FirstBLG20005,Ratttenbury2006PLANETARYDISCOVERY}. Nevertheless, it took almost a decade for the first confirmed exoplanet by a microlensing observation to be published by~\cite{Bond2004OGLEEvent}.

Currently there are over 80 confirmed planets discoveries through microlensing observations\footnote{NASA Exoplanet Archive, 2020. \url{https://exoplanetarchive.ipac.caltech.edu/docs/counts_detail.html}}.
%~\citep[e.g.][etc]{Dong2006PlanetaryOGLE2004BLG343,Hwang2017OGLE-2017-BLG-0173Lb:Event,Jung_2017,Miyazaki2018MOA-2015-BLG-337:Binary}.
The rate of detections has increased in recent years with the advent of the Korean Microlensing
Telescope Network (KMTNet)~\citep{Kim2016KMTNet:Observatories}.
% ~{\footnote{KMTNet is a project fully
% funded by the Korean government.}}

This work addresses the microlensing event {OGLE-2018-BLG-0677.} This is a relatively-faint, moderate-magnification
event with some evidence for an anomaly soon after peak brightness.
From a comparison with single-lens, binary-lens and { binary-source models,} we will show
a strong preference for a binary lens interpretation. As we can only obtain directly 
two physical quantities for the {system we perform} a Bayesian 
analysis to infer the probable physical properties for the host (lens) and companion. 
The resulting distributions suggest that the companion is a low-mass planet.

In Section 2 of this paper we briefly describe the observations of the event. 
Section 3 describes our  model selection, as well as the details of
the light curve fitting process, and presents the best-fit values. {In Section 4, we investigate 
the angular size of the source and its implications for the Einstein radius, and in 
Section 5 we give a detailed description of the approach taken for the Bayesian analysis used to 
infer the properties of the lensing system, and present the resulting distributions.
From these we derive the planet mass and the separation from its host. In Section 6 we discuss some of the implications of this work and, finally, Section 7 is a brief discussion concluding the paper. } 

\section{OBSERVATIONS} \label{sec:kmtnetobs}

\subsection{OGLE}

The event was first detected
by the Optical Gravitational Lensing Experiment {(OGLE) Early Warning System (EWS, \citealt{ews1,ews2}),} with designations OGLE-2018-BLG-0677 and 
OGLE-2018-BLG-0680, since it lies in the overlap region of two survey fields.
It is located at
$(RA,Dec)=(17^h55^m00^s.27,-32^{\circ}00'59".51)${ , which} 
corresponds to $(l,b) = (-1^{\circ}.61, -3^{\circ}.31)$ 
in Galactic coordinates. 
{ In combination, the observations }
from the two fields have a frequency of 1-3 data points per day. 
The OGLE observations {were reduced using difference image analysis (DIA) from \citet{Wozniak:2000tz}.}

\subsection{KMTNet \label{subsec:obsblg0816}}
The Korea Microlensing Telescope Network (KMTNet)  is a wide-field imaging system, with three 
telescopes and cameras sharing the same specifications, installed at { Cerro-Tololo Inter-American Observatory in Chile~(KMTC),
the South African Astronomical Observatory in South Africa~(KMTS), and the Siding Spring 
Observatory in Australia~(KMTA).} The telescopes each have a 1.6 m primary mirror, and a wide-field 
camera (a mosaic of four $9k \times 9k$ CCDs) that image approximately a  $2.0 \times 2.0$ square degree { field of view.}

Weather permitting, the network  of telescopes and cameras allow a 24 hour per day monitoring of the Galactic Bulge.
{ This allows one to} trace the light curves of stars continuously and is ideal for the detection of
extrasolar planets by microlensing and transit, variable objects, and asteroids and comets.

The event OGLE-2018-BLG-0677 was independently detected by KMTNet and given the designation KMT-2018-BLG-0816~\citep{kmtnetEF}. The event is located in the BLG01 and BLG41 KMTNet fields,
giving an effective observation cadence of 15 minutes. 

\begin{figure}
\includegraphics[width = 0.99\columnwidth]{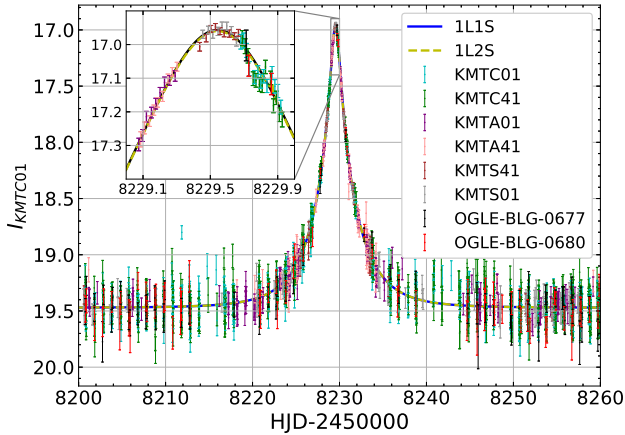}
%\plotone{Data}
\caption{Light curve for OGLE-2018-BLG-0677 fitted with 1L1S and 1L2S models (which overlap and cannot be 
 distinguished visually). The small window zoomed into the section of the light curve presents a small { anomaly, which} cannot
 be accounted for by the models.
\label{fig:data}}
\end{figure}

Photometry was extracted from the KMTNet observations using the software package PYDIA~\citep{MichaelDAlbrow2017MichaelDAlbrow/pyDIA:Github.}, which employs a difference-imaging algorithm based on the modified- delta-basis-function approach of \cite{Bramich2013DifferenceConsiderations}. 
The light curve of the event, with a single lens single source~(1L1S) model, is shown in
Fig.~\ref{fig:data}. 

\section{LIGHT CURVE ANALYSIS} \label{sec:lightcurve}

The following sections present the several models fitted to the data as well as the reasoning by which we selected the best among them.

\subsection{Single lens single source} \label{sec:1S1L}

This is the simplest of  models, which considers a point mass lens with a point mass source. The magnification is modeled
by a {\citet{Paczynski1986GravitationalHalo} curve, }
\begin{equation}
    A(u) =  \frac{u^2+2}{u\sqrt{u^2+4}},
\end{equation}
where $u$ is { the angular separation between source and lens }normalized by the Einstein { angle $\theta_{\rm E}$. Given} the relative motion between them, this separation will be a function of time and is assumed rectilinear as
\begin{equation}
    u(t) = (\tau^2+u^2_0)^{1/2}\, ,
\end{equation}
with $\tau\equiv (t-t_0)/t_{\rm E}$; $u_0$ is the impact parameter of the event, $t_0$ is the time at $u=u_0$, and $t_{\rm E}$ is the { Einstein radius crossing time.} These { three} parameters characterize completely the { light-curve magnification model,} $A(t)$. We find these parameters by a Markov Chain Monte Carlo (MCMC) search while reducing the $\chi^2$ of a linear fit to the observed flux, $F(t) = A(t)F_{\rm S} + F_{\rm B}$, where the source and blend flux, $F_{\rm S}$ and $F_{\rm B}$, are determined for each data set.

The best-fit values (after renormalization of data uncertainties as will be discussed in Section~\ref{subsec:lightfit}) 
are presented in Table~\ref{tab:param1s1l}. { We note that in Tables~\ref{tab:param1s1l}-\ref{tab:param},
$F_{\text{S}}$ and $F_{\text{B}}$ are given in a system with 18 as the magnitude zero point.}

\begin{deluxetable}{c c}[H]
\tablecaption{Best-Fit 1L1S Model parameters.
\label{tab:param1s1l}}
\tablehead{}
\startdata
$\chi^2_{min}/N_{\text{data}} $ &$ 1602.51/1557$ \\
$u_0$ & $0.1029\pm 0.0012$ \\
$t_0$ & $8229.5417 \pm 0.0007$ \\
$t_{\rm E}$ (days) & $4.95 \pm 0.04$ \\
$F_{\text{S,OGLE0677}}$ & $0.293 \pm 0.009 $ \\
$F_{\text{B,OGLE0677}}$ & $ -0.065 \pm 0.011 $ \\
\enddata
\end{deluxetable}

\subsection{Anomaly}

As can be seen in Figure~\ref{fig:data}, there is a small anomalous feature in the light curve relative to the 1L1S model that occurs over 
{$\sim$ 5 hr during the interval 8229.70 - 8229.90. The anomaly primarily takes the form of a dip of $\sim 0.05$ mag followed by a smaller and shorter bump.  See the residuals in Figure~\ref{fig:single}.  These features are well traced by the KMTC01 and KMTC41 data sets, and they are confirmed by two points from the OGLE-2018-BLG-0680 data, one each on the dip and the subsequent small bump.} The OGLE-2018-BLG-0677 data set
has a single point just before the start of the anomaly. We have examined the direct and difference images from { KMTC} and are 
satisfied that there are no systematic effects that can be attributed to seeing, background, or image cosmetics that could cause these features.
{
Additionally, we have found no evidence that the dip is a repeating phenomenon, such as might be due to a star spot.
}

\subsection{Single lens binary source}\label{sec:1L2S}

{ We examine the possibility} that the anomaly is due to a binary source.
In the case that the source consists of two stars, the total flux is given by the linear
combination of the  individual source fluxes, $F_{\rm tot} = A_1F_1+A_2F_2 + F_{\rm B}$, where 
$A_1$ and $A_2$ are the individual magnifications of each source~\citep{Gaudi1998DistinguishingPerturbations} as { parameterized in Section~\ref{sec:1S1L},} but sharing the same value of $t_{\rm E}$. 

\begin{figure}
\includegraphics[width = 0.99\columnwidth]{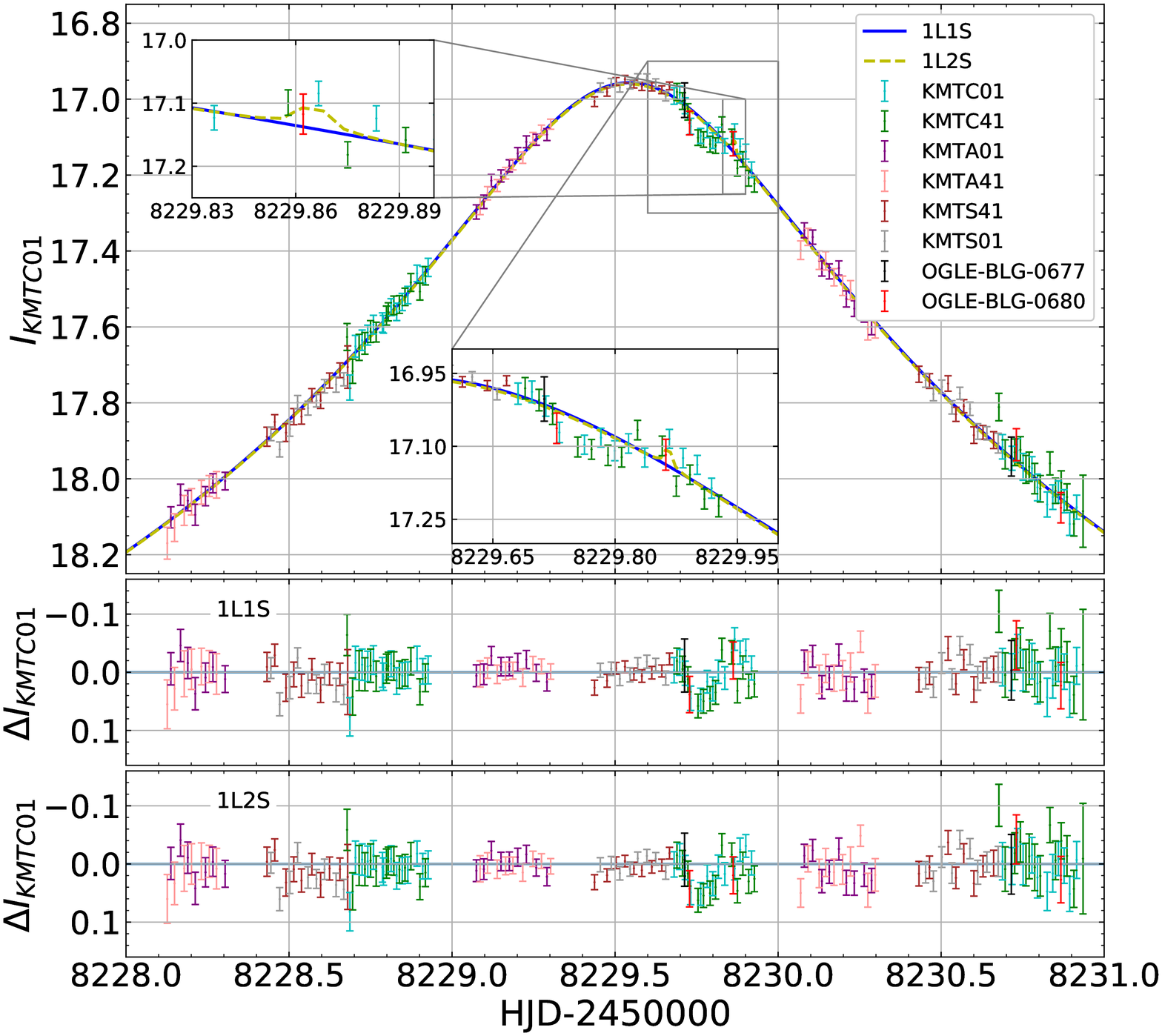}
%\plotone{Both}
\caption{{Light curve for OGLE-2018-BLG-0677, together with 1L1S (solid line) and 1L2S (dashed line) (upper panel) and residuals from these models (lower two panels).
It is clear the two solutions overlap and predict essentially the same light curve, except for an interval of 0.5 hours
near 8229.86} \label{fig:single}}
\end{figure}

The total magnification of the combined source flux is given by
\begin{equation}
    A_{\rm tot} = \frac{A_1+A_2q_F}{1+q_F}\, ,
    \label{eq:binamp}
\end{equation}
where $q_F=F_2/F_1$ is the luminosity ratio~\citep{Griest1992EffectMicrolensing}. In total, there are 6 parameters: $u_{01}$ and $t_{01}$ for $A_1$, $u_{02}$ and $t_{02}$  for $A_2$, $t_{\rm E}$ is shared by $A_1$ and $A_2$;\, and the luminosity ratio $q_F$. We fit the data as described in the previous subsection
but using this model instead. The best-fit values found for the data are presented in Table~\ref{tab:param2s1l},
and the { light-curve model shown in Figure~\ref{fig:single}.}
{It is apparent that the best binary source model does not reproduce the anomalous feature in the light curve, but produces a small ``bump" deviating almost imperceptibly from the 1L1S model.}

\begin{deluxetable}{c c}[H]
\tablecaption{Best-Fit 1L2S Model parameters\label{tab:param2s1l}}
\tablehead{}
\startdata
$\chi^2_{min}/N_{\text{data}} $ &$ 1593.08/1557$ \\
$u_{01}$ & $0.103 \pm 0.004$ \\
$t_{01}$ & $8229.542 \pm 0.004$ \\
$u_{02}$ & ${0.05}^{+0.31}_{-0.05}$
 \\
$t_{02}$ & ${8229.873}^{+0.546}_{-0.061}$\\
$q_F$ & ${0.0031}^{+0.0317}_{-0.0028}$\\
$t_{\rm E}$ (days) & $4.96\pm 0.14$ \\
$F_{\text{S,OGLE0677}}$ & $ 0.287 \pm 0.009 $ \\
$F_{\text{B,OGLE0677}}$ & $ -0.063 \pm 0.011 $ \\
\enddata
\end{deluxetable}

\subsection{Binary lens}\label{subsec:lightfit}

We adopt the standard parameterization of a binary lens light curve by describing it with 7 
parameters~\citep{Gaudi2012MicrolensingExoplanets}:
$s$, $q$, $\rho_\ast$, $\alpha$, $u_0$, $t_0$ and $t_{\rm E}$. These represent the 
normalized separation between { binary-lens components,} the mass ratio of the binary-lens components, 
the normalized source radius, the source-trajectory {angle with respect to the binary} axis, the 
normalized closest approach {between the lens center of mass} and the source (which occurs at 
time $t_0$, the time of closest approach), and the timescale to cross the Einstein
radius, respectively. The factor used for the normalized parameters is the angular Einstein radius, 
\begin{equation}
\theta_{\rm E} = \sqrt{ \frac{4GM_L}{c^2} \left( \frac{1}{D_L } - \frac{1}{D_S} \right) }, 
\end{equation}
where $M_L$ is the total mass of the lens system,  and $D_L$, $D_S$ are the distances from Earth to the lens and source.
A visualization of this combination of parameters can be found
in~\cite{Jung2015OGLE-2013-BLG-0102LAB:BOUNDARIES}. 

The fitting is done by a Maximum Likelihood Estimation, which is equivalent to
minimization of the $\chi^2$. The process is done in two parts. The first is through 
a fixed value grid search of the $(s,q)$ parameters to find regions where the minimum $\chi^2$ 
may be located. For each $(s,q)$, a grid of $(r, \alpha)$ (where $r$ is a reparameterization of $u_0$, centred-on and
normalized to caustics,
see~\cite{McDougall2016MicrolensingGPUs}) is used to seed a minimization over ($\rho_\ast, t_0, t_{\rm E})$ by a simple Nelder-Mead optimization~\citep{Nelder-mead}.
This approach for the fixed $(s,q, r, \alpha)$ is due to the relation of these parameters to the geometry.

Our fixed position results are then used as seeds for a more refined search using a Markov Chain Monte Carlo~(MCMC)
algorithm implemented by using the MORSE code~\citep{McDougall2016MicrolensingGPUs}.
A similar two step process can also { be seen in~\cite{Shin2019Two2017}.}
During this second part of the process, the seed solutions from the
grid search are used as starting points, and the search is now 
continuous in the parameter space. 

Once a minimum value for $\chi^2$
was found, the original magnitude uncertainties for each data set were renormalized via
\begin{equation}
\sigma'_i = k \sqrt{ \sigma_i^2 + \epsilon^2}.
\end{equation}
The coefficients $k$ and $\epsilon$ for each data set are determined in such a way that the reduced $\chi^2$ for { both the} higher- and lower-magnification data points for each site are approximately unity \citep{Yee2012MOA-2011-BLG-293Lb:DETECTIONS}. 
The renormalization factors are listed in Table~\ref{tab:renorm}.
The MCMC search was then re-run with the renormalized data uncertainties to corroborate the solution and obtain a 
new minimized $\chi^2$. The final model parameter values and their uncertainties are
directly obtained from the 68\% confidence interval around the medians of the
marginalized posterior parameter distributions. 

Given the short total time of the event, { it was not possible to obtain
parallax information~\citep{Gould1992ExtendingMasses, Alcock1995FirstEvent, Gould2000AMicrolensing, Buchalter2002RatesBulge}. }
For the rest of our { analysis, no higher order light-curve effects} were considered.

\begin{figure}
\includegraphics[width = 0.99\columnwidth]{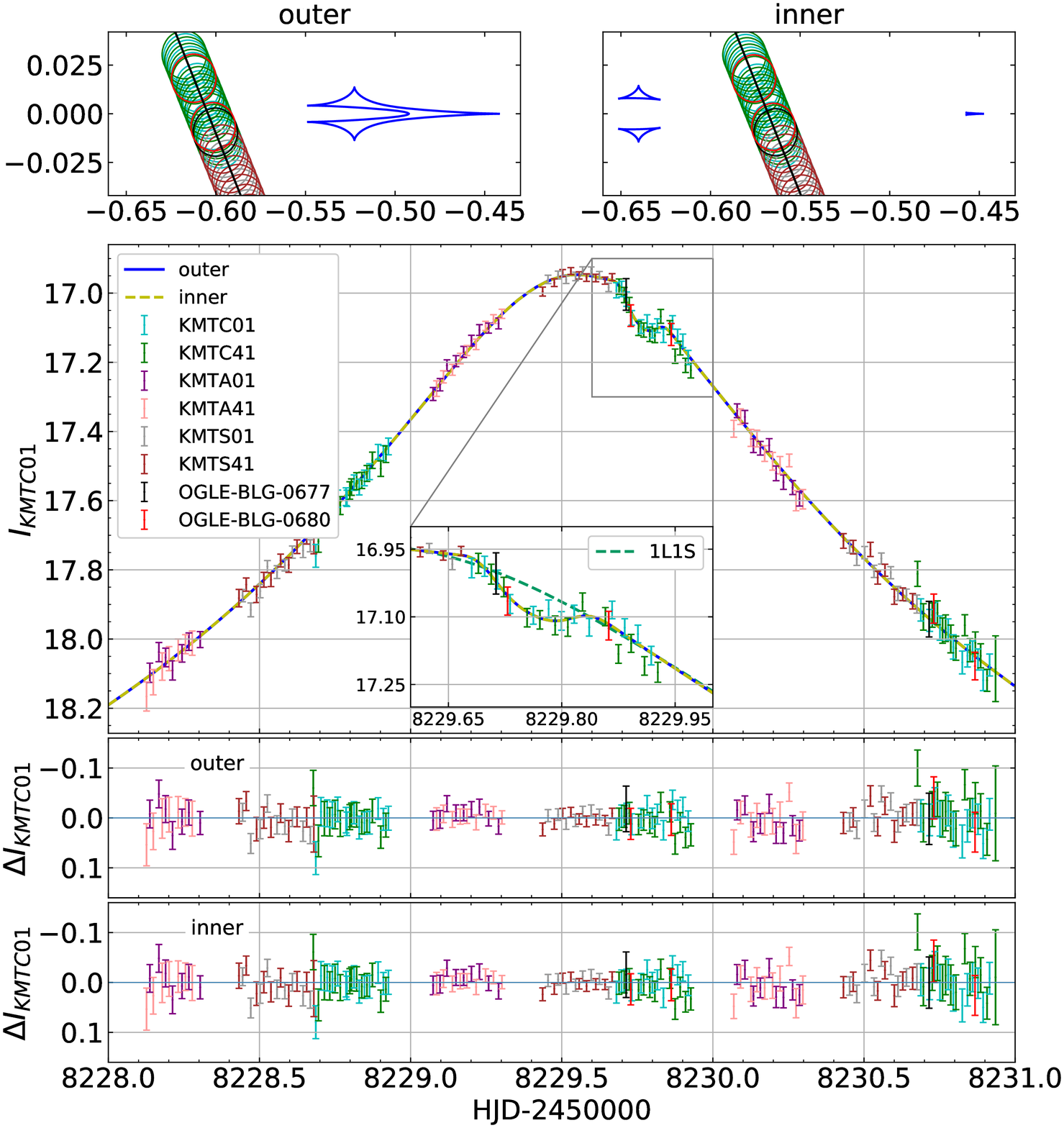}
%\plotone{Both}
\caption{Best-fit models for binary {lens and corresponding caustic geometries.} The upper panels show the
two possible geometries of inner and outer solution. The remaining panels show the 
observed light curve for both solutions, which are indistinguishable, and 
their respective residuals.\label{fig:both}}
\end{figure}

\subsubsection{Model of OGLE-2018-BLG-0677}\label{subsec:model}

{
From the fitting process described in the previous section, 
we found two degenerate solutions with similar $q$. Often this situation corresponds to
a well-known wide/close case degeneracy where $s < 1$ and $s > 1$ for the two solutions 
~\citep[see e.g.,][]{Griest1998,Dominik1999TheCases,Batista2011MOA-2009-BLG-387Lb:Dwarf}. 
However, in the present case, the degeneracy is between two solutions with $s < 1$,
and is due to a source-trajectory passage past two cusps of connected or disconnected 
caustics, as shown in Figure~\ref{fig:both}.
Each corresponds to a minor-image perturbation, that demagnifies the source relative to
the magnification due solely to the host star.
This type of degeneracy has been detected previously for events 
OGLE-2012-BLG-0950~\citep{koshimoto2017} and
OGLE-2016-BLG-1067~\citep{novati2019}. A thorough explanation of the phenomenon
is given in \citet{Han_2018} with reference to the event MOA-2016-BLG-319. 
Following the nomenclature of that paper, we refer to the case where the source passes 
between the caustics as the "inner" solution, and that where it passes the single connected
caustic as the "outer" solution.
}

For the current event, OGLE-2018-BLG-0677, we could not break this degeneracy
given that both cases {produce  almost identical light curves,}  see Figure~\ref{fig:both}. 
Table~\ref{tab:param}
lists the best-fit parameters for both solutions. Apart from the clear
difference in geometry, they share several parameters, $q$, $u_0$,
$t_0$, $t_{\rm E}$, well within their respective uncertainties. Also, the values of 
the minimum $\chi^2$ are close enough that there is no clear statistical
difference. Despite this degeneracy, { the similarities of $q$, $\rho$ and $t_{\rm E}$ will} allow us
to infer similar physical properties for the system, which will be discussed 
in the following section. 

As a note, $\rho_\ast$ is detected rather weakly in the light-curve analysis. It may be of interest to ask whether
the omission of $\rho_\ast$ affects our subsequent results in any significant way.

Therefore in section~\ref{properties} we give results
that propagate the $\rho_\ast$ distribution, and also show the
the case where the $\rho_\ast$ information is {omitted from the Bayesian estimates.}

\begin{deluxetable}{c c c}[H]
\tablecaption{Best-Fit Binary Model parameters\label{tab:param} 
}
\tablehead{\colhead{} & \colhead{inner} & \colhead{outer} }
\startdata
$\chi^2_{min}/N_{\text{data}} $ &$ 1556.03/1557$ & $1555.96/1557$ \\
$ s $ & ${0.912}^{+0.002}_{-0.054}$ & ${0.985}^{+0.059}_{-0.002}$ \\
$\log_{10} q$ & ${-4.105}^{+0.305}_{-0.0822}$ & ${-4.054}^{+0.268}_{-0.109}$\\
$\rho_\ast$ &  ${0.01209}^{+0.00013}_{-0.00545}$ & ${0.01238}^{+0.00016}_{-0.00590}$ \\
$u_0$ & $0.102 \pm 0.003$ & $0.102 \pm 0.003$  \\
$\alpha$ & $-1.98\pm 0.01$ & $4.307\pm 0.008$ \\
$t_0$ & $8229.544\pm 0.002$ &  $8229.544\pm 0.002$ \\
$t_{\rm E}$ (days) & $4.94\pm 0.11$ & $4.94\pm 0.11$ \\
$F_{\text{S,OGLE0677}}$ & $0.294 \pm 0.009 $ & $0.294 \pm 0.009 $ \\
$F_{\text{B,OGLE0677}}$  & $-0.069 \pm 0.011 $ & $-0.070 \pm 0.011 $ \\
\enddata
\end{deluxetable}
% old
% \begin{deluxetable}{c c c}[H]
% \tablecaption{Best-Fit Binary Model parameters\label{tab:param} 
% }
% \tablehead{\colhead{} & \colhead{close} & \colhead{wide} }
% \startdata
% $\chi^2_{min}/dof $ &$ 1226.30/1422$ & $1226.08/1422$ \\
% $ s $ & $0.855\pm{0.009}$ & $ 1.045\pm{0.015}$ \\
% $\log_{10} q$ & $-3.82\pm 0.10$ & $ -3.83\pm 0.10$ \\
% $\rho_\ast$ & $ - < 0.011$ & $ - < 0.011$ \\
% $u_0$ & $0.102 \pm 0.003$ & $0.102 \pm 0.003$  \\
% $\alpha$ & $-1.98\pm 0.01$ & $4.307\pm 0.008$ \\
% $t_0$ & $8229.544\pm 0.002$ &  $8229.544\pm 0.002$ \\
% $t_{\rm E}$ (days) & $4.93\pm 0.13$ & $4.93\pm 0.13$ \\
% %%$F_{\text{S,CTIOI01}}$ & $171.06$ & $171.15$ \\
% %%$F_{\text{B,CTIOI01}}$  & $-8.29$ & $-8.43$ \\
% \enddata
% \end{deluxetable}

\subsubsection{Other local minima}\label{subsec:local}

\begin{figure}
\includegraphics[width = 0.99\columnwidth]{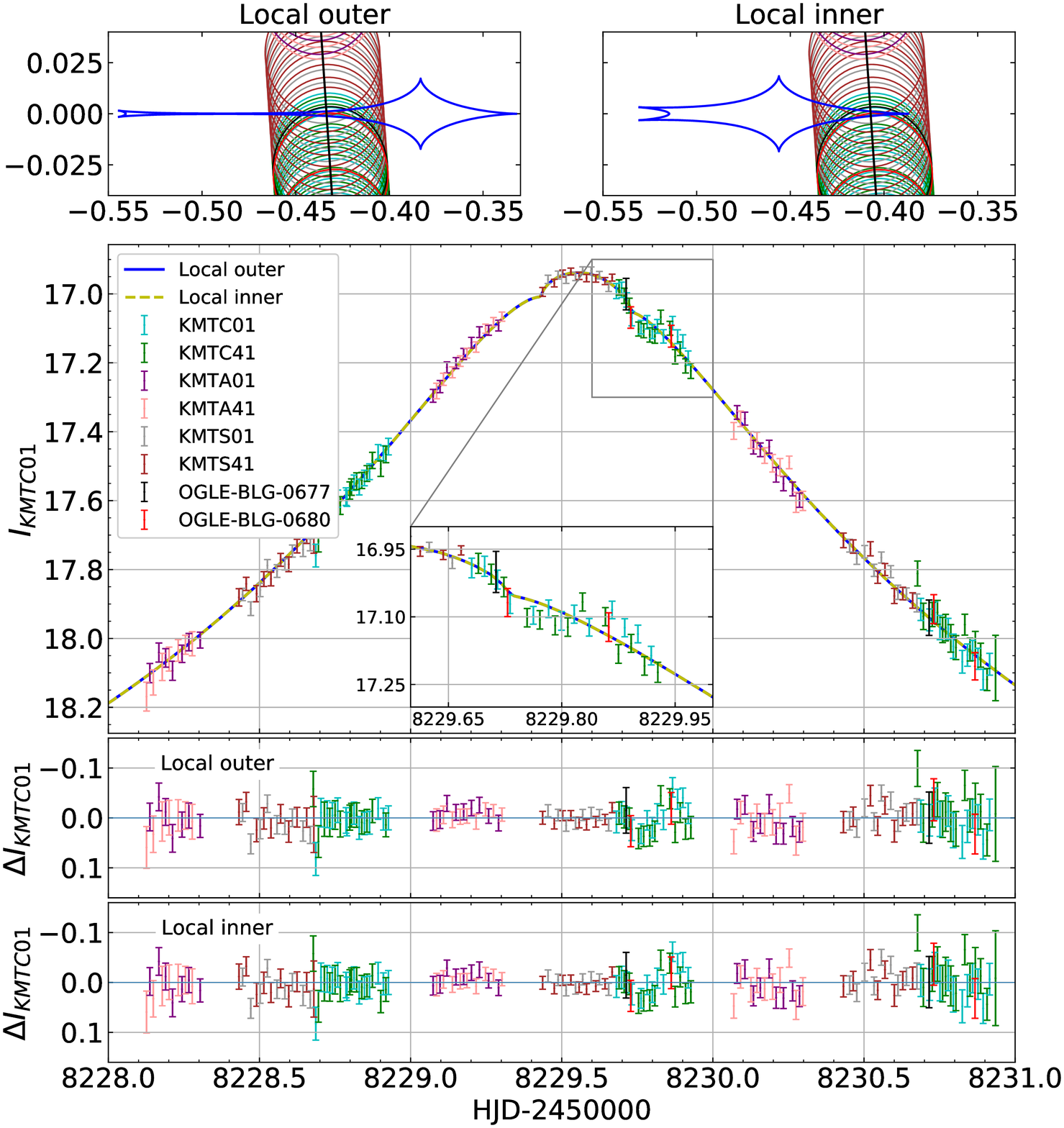}
\caption{Light-curve fit and geometries for the local inner and outer minimum models for binary lens. These two solutions are degenerate with each other, but disfavored relative to the best-fit models shown in Figure 3.\label{fig:localminima}}
\end{figure}

{
Five additional seed solutions were identified in our initial grid search close, located close in $(s,q)$ to the solutions reported above.
All of these are very shallow in $\chi^2$-space, and with subsequent MCMC runs, all converged to one of either of
the reported solutions. 
Light curves and caustic geometries for the two most prominent of these local minima are shown in Figure~\ref{fig:localminima}.
}

\begin{deluxetable}{c c c}[H]
\tablecaption{Local Minima Binary Model parameters\label{tab:localparam} 
}
\tablehead{\colhead{} & \colhead{inner} & \colhead{outer} }
\startdata
$\chi^2_{min}/N_{\text{data}} $ & $1580.19/1557$ &  $ 1579.82/1557$\\
$ s $ &$1.030$  & $1.082$ \\
$\log_{10} q$ & $-3.732$ &  ${-3.725}$ \\
$\rho_\ast$ &  $0.031$ & $0.031$ \\
$u_0$ & $0.107$ & $0.107$  \\
$\alpha$ & $1.498$ & $1.498$ \\
$t_0$ & $8229.543$ &  $8229.543$ \\
$t_{\rm E}$ (days) & $4.837$  & $4.852$\\
$F_{\text{S,OGLE0677}}$ & $0.279 \pm 0.001 $ &  $0.276 \pm 0.001 $\\
$F_{\text{B,OGLE0677}}$  & $-0.020 \pm 0.002 $ &  $-0.018 \pm 0.002 $\\
\enddata
\end{deluxetable}

\begin{figure}
\includegraphics[width = 0.99\columnwidth]{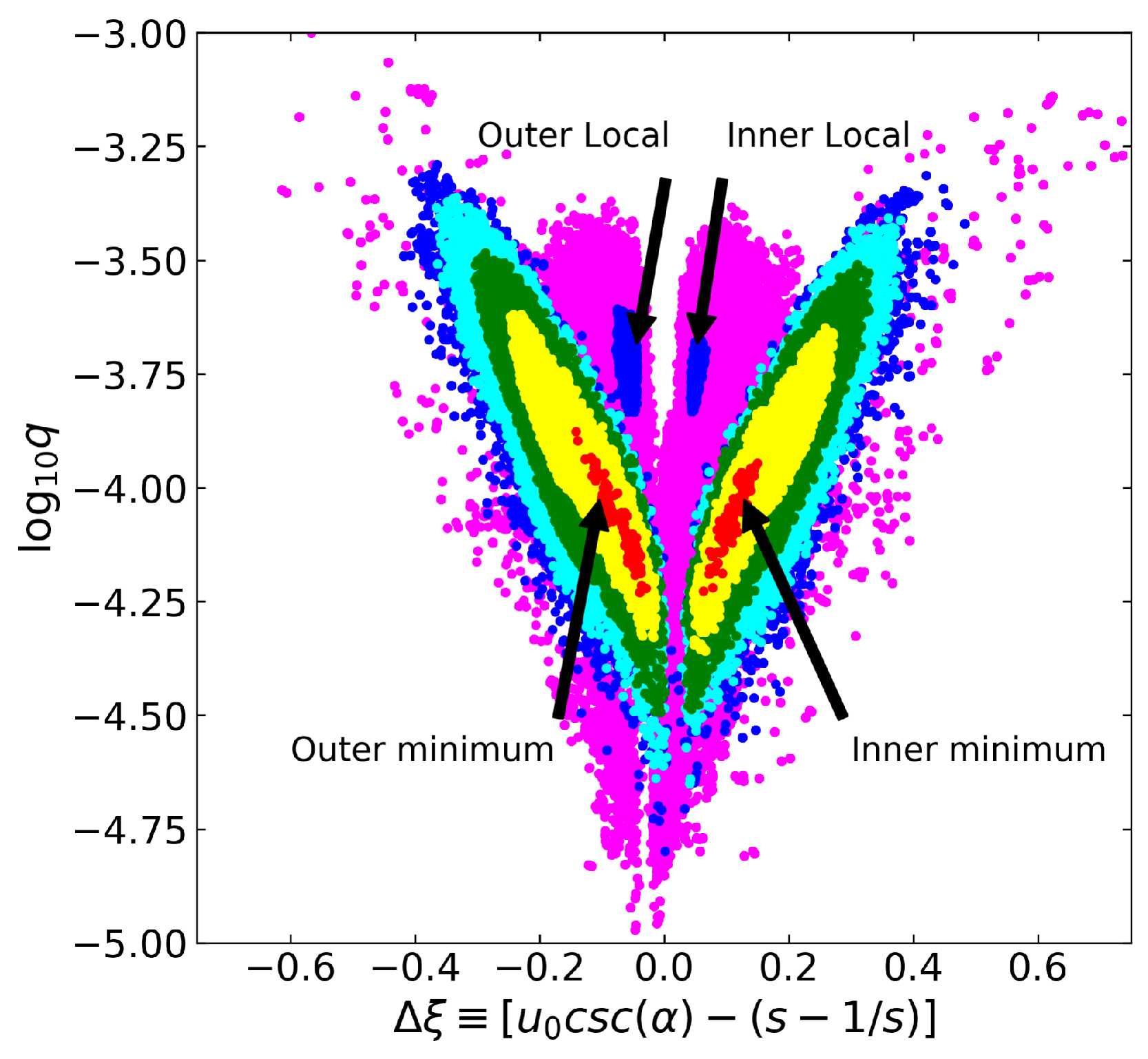}
\caption{Scatter plot of $\Delta \xi$ vs. $\log_{10} q$, where $\Delta \xi$ is the offset between the center of the source and the caustic as the source crosses the planet-star axis. Color coding  is red, yellow, green, cyan, blue, magenta for $\Delta\chi^2 < 1, 4, 9, 16 ,25 ,36$. Values of $\Delta\chi^2 > 36$ are omitted given the closeness to the 1L1S model, $\Delta\chi^2 \sim 10$, and also to help the readability of the scatter plot.
\label{fig:hotcold}}
\end{figure}

{
To analyze how these local minima are related to the best-fit 
solutions, we followed the procedure 
described in~\cite{Hwang2017OGLE-2017-BLG-0173Lb:Event} and ran various ``hot" and ``cold" MCMC realisations to explore the regions around all the solutions. In Figure~\ref{fig:hotcold} we present a scatter plot of the combined samples from the various runs, color coded by their $\Delta \chi^2$ relative to the best solutions.
We adopt the parameter
\begin{equation}
    \Delta \xi = u_0 \csc \alpha - (s-1)/s,
\end{equation}
introduced by \cite{Hwang2017OGLE-2017-BLG-0173Lb:Event}, which traces the 
offset between the centers of the
source and the major-image caustic as the source crosses the planet - star axis.
Three of the identified extra solutions are located within the purple region below the best-fit solutions, with $25 < \Delta \chi^2 < 36$, and are indistinguishable from their surroundings.
The two remaining local minima, i.e. those displayed in Figure~\ref{fig:localminima} and 
labelled as outer and inner local minima in Figure~\ref{fig:hotcold},
have $16 < \Delta \chi^2 < 25$. They are degenerate, and correspond to a lens mass ratios $~\sim$ twice that of the best solutions. They each represent a source trajectory over an on-axis cusp that provides a single bump in magnification, 
{
and correspond
to a major-image perturbation with degeneracies as discussed in \citet{gaudi1997b}.
}

Their light curve fit  parameters are listed in Table~\ref{tab:localparam}, without uncertainties as they are too shallow in $\chi^2$ space to retain an MCMC chain. They have a similar $t_E$ to the best solutions, but a larger $\rho_*$. 
}

\subsection{Model selection}\label{subsec:lensmod}
From Figure~\ref{fig:data}, it can be seen that there is a subtle anomaly following the peak of the light curve
compared {to the 1L1S model.} Although a binary lens model could in principle represent this small feature found 
in the data, the justification for the increase in model complexity needs to be strengthened; therefore,  
to avoid overfitting by assuming a binary lens~\citep{Gaudi1997PlanetarySources},
we compared with two simpler models. The three models compared are the previously-described single lens single source~(1L1S), 
single lens binary source~(1L2S) and binary lens single source~(2L1S) models.
Often the $\chi^2$ value is used as an indicator of the goodness-of-fit, but this time
instead of simply relying on this statistic to compare between different models, 
we obtain the log evidence given the data, $\ln \mathcal{Z}$, for a more robust model comparison.
This was done by applying the Nested Sampling method~\citep{Skilling2006NestedComputation}
for cases { (1L1S) and (1L2S).} For the case of (2L1S), nested-sampling was too inefficient, so the 
approximation algorithm presented in~\cite{vanHaasteren2009BayesianMethods} for
posterior distributions was used instead to estimate $\ln \mathcal{Z}$. For completeness,
we also applied the approximate algorithm to the other { two cases, which led} to the same results as nested sampling. 
The log odds ratio is obtained by 
$\ln O = \ln \mathcal{Z}_{M1} - \ln \mathcal{Z}_{M2}$, and a model is preferred as long
as $\ln O > 0$, but a strong preference is given when 
$\ln O > 5$~\citep{sivia2006data,jaynes2003probability}.

{
The 2L1S best-fit has two degenerate
solutions, inner and
outer~\citep[see][]{Griest1998}, but the difference in $\chi^2$ 
between these ($\Delta \chi^2 = 0.07$) is
small, with a slight preference to the outer model. On the other
hand, the log evidence gives a strong preference for
the inner model. We need to adopt a model for 
renormalization, so we decided to choose the inner solution
for this purpose.
}

Table~\ref{tab:evi} shows the the minimum $\chi^2$ and the log evidence based on the renormalized data error bars, 
and it is clear that
the 2L1S solution is preferred with $\Delta\chi^2 > 46$  compared with the 1L1S, and is also preferred from the evidence.

One last consideration worth mentioning is the case for 1L2S. From the $\chi^2$ point of view it seems to improve compared to 1L1S, however the evidence gives an indication that the 1L1S is slightly better.
Here the meaning of the evidence becomes clear. It says that the 1L2S model is not better than the 1L1S, 
(the extra model complexity does not outweigh the reduction in { $\chi^2$),
and it} in fact the difference with the 1L1S model is only between 8229.85 and 8229.88. 
This can be clearly seen in { Figure~\ref{fig:single}} with the 1L1S and 1L2S ``best-fit" light curves.

\begin{deluxetable}{c c c c c }
\tablecaption{Minimum $\chi^2$ and log evidence {for 1557 data points} for the 
different lensing models.
\label{tab:evi}}
\tablehead{\colhead{Model} & \colhead{1L1S} & \colhead{1L2S} & \colhead{2L1S(inner)} & \colhead{2L1S(outer)}}
\startdata
$\chi^2_{\rm renorm}$ & 1602.51 & 1593.08 & 1556.03 & 1555.96 \\
$\ln \mathcal{Z}_{\rm renorm}$ & -811.33 & -815.11 &  -806.15 & -807.23 \\
\hline
$\Delta \chi^2_{\rm renorm}$ & 0.00 & 9.43 & 46.48 & 46.55 \\
$\Delta \ln \mathcal{Z}_{\rm renorm}$ & 0.0 & -4.29 & 5.18 & 4.1 \\
\enddata
\end{deluxetable}

\section{COLOR MAGNITUDE DIAGRAM}

\begin{figure}
 \includegraphics[width = 0.99\columnwidth]{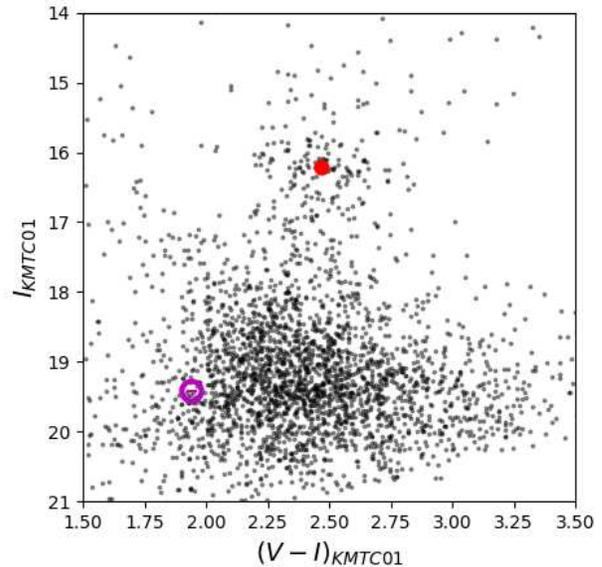}
\caption{Color magnitude diagram of the field from {the KMTC01 images}. The red clump centroid is
shown as a red dot, and the source, inferred from the { light-curve model} fit,
is shown as (very small) magenta error bars, with a magenta circle to highlight its position.}
\label{fig:CMD}
\end{figure}

A sample instrumental color magnitude diagram (CMD) for a 1.5 x 1.5 arcmin field around OGLE-2018-BLG-0677 
is shown in Figure~\ref{fig:CMD} based on { KMTC data}. The source position is indicated,
with its magnitude determined from the source flux inferred by the { light-curve model,} and its color from
a regression of { $V$-band} difference flux against { $I$-band} difference flux.

From an analysis of four such CMDs { (KMTC01, KMTC41, KMTS01, and KMTS41),} we found a color
offset from the red clump, { $(V-I)_s - (V-I)_{\rm RC} = -0.47 \pm 0.04$,} and a magnitude offset,
{ $I_s - I_{\rm RC} = 3.20 \pm 0.03$.} Combining these with the red clump intrinsic color,
$(V-I)_{\rm RC,0} = 1.06$ \citep{bensby2013} and magnitude for this Galactic longitude,
$I_{\rm RC,0} = 14.534$ \citep{nataf2013}, we find for the source that
{ $(V-I)_{0,s} = 0.59 \pm 0.04$ and $I_{0,s} = 17.73 \pm 0.03$.}

From \citet{bessell1988} we convert our color to $(V-K)_0 = 1.25 \pm 0.10$, and 
using the surface-brightness relations of \citet{kervella2004} we determine the
source { angular radius $\theta_\ast = 0.789 \pm 0.025$ $\mu$as.}

% It should be noticed from the binary light curve fitting results the lack of a fitted value for 
% $\rho_\ast = \theta_\ast/\theta_{\rm E} $.
% This is due to the lack of finite source effects for both solutions,
% and implies that it is not possible to estimate
% the angular Einstein radius, $\theta_{\rm E}$, 
% from the combination of the light-curve
% and color-magnitude diagram~\citep{Yoo2004ConstraintsOGLE2003BLG423}. 
% We set a maximum value for $\rho_{\ast \text{max}} = 0.02$, and with the value of
% $\theta_\ast = 0.789\,\mu$as obtained from the color-magnitude diagram, we set for the galactic modeling to come below in Section~\ref{propr},  a threshold where the system must have $\theta_{E} > 71.72\,\mu$as.

\begin{deluxetable}{c c c }[H]
\tablecaption{Renormalization values. \label{tab:renorm}}
\tablehead{ \colhead{inner/outer} & \colhead{scale factor}
& \colhead{added uncertainty}}
\startdata
KMTC01	& 2.543	& 0.003	\\
KMTC41	&	3.038 &	0.006\\	
KMTA01	&	1.904 &	0.006\\	
KMTA41	&	3.04 & 0.000 \\
KMTS01	&	1.956 &	0.005 \\	
KMTS41	&	1.828 &	0.003\\
OGLE-BLG-0677 & 1.429 & 0.031 \\
OGLE-BLG-0680 &	1.463 & 0.020 \\
\enddata
\end{deluxetable}

\begin{figure}[H]
 \includegraphics[width = 0.99\columnwidth]{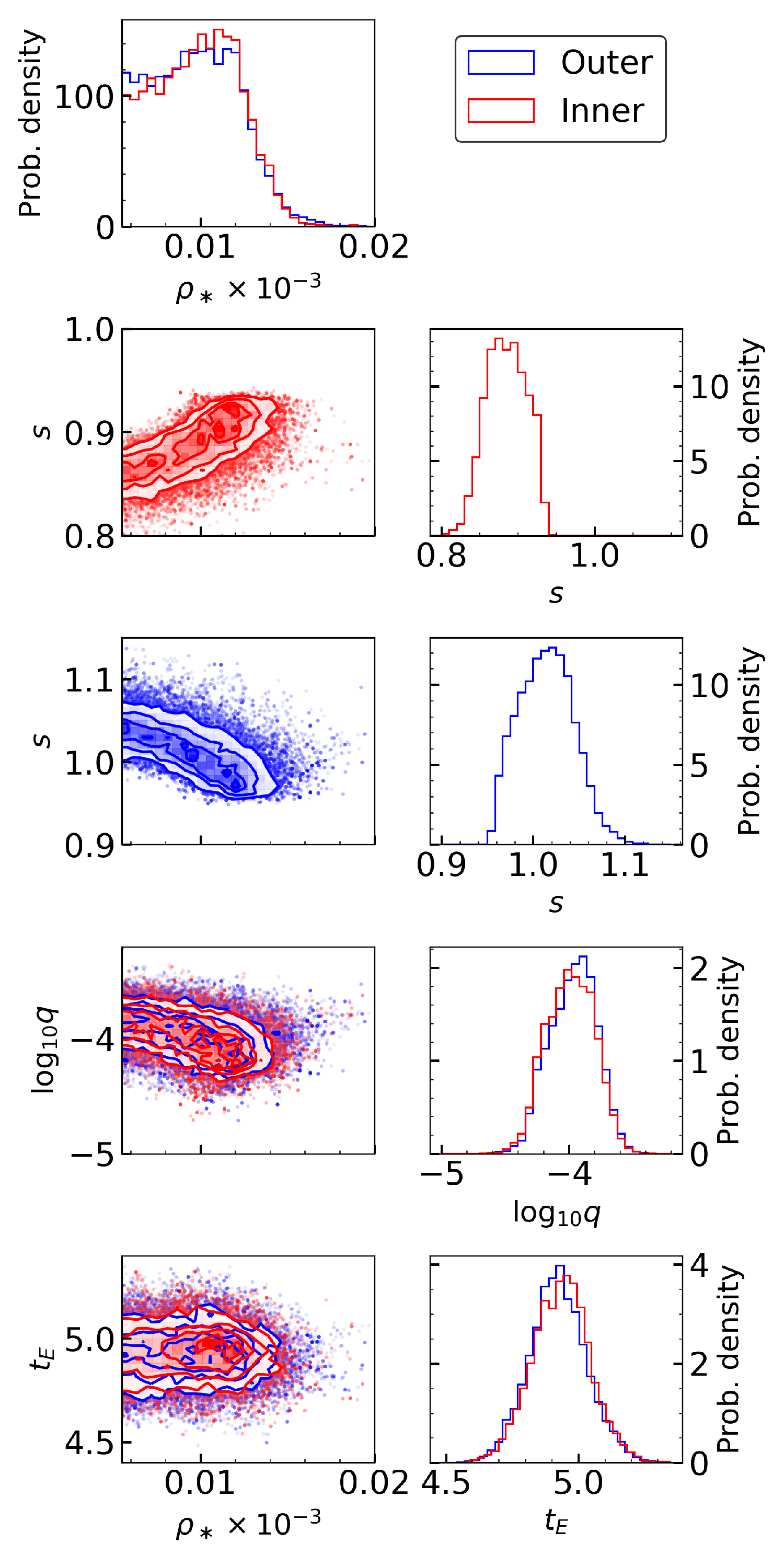}
%\plotone{probte}
\caption{MCMC chains and posterior distributions. The first left {panel displays the fully-marginalized} distributions for $\rho_\ast$, which shows how the inner and outer distributions overlap. The right panels show the
fully-marginalized distributions for $s$, $log_{10}q$ and $t_E$ for the outer (blue) and inner (red)
solutions, which will be used to derive
the {properties of the host and companion.} The four left lower panels show the Markov chain points for these parameters projected against $\rho_\ast$. 
\label{fig:te}}
\end{figure}

%old
% \begin{deluxetable}{c c c c c }
% \tablecaption{Minimum $\chi^2$ and log evidence for 1444 data points for the 
% different lensing models.
% \label{tab:evi}}
% \tablehead{\colhead{Model} & \colhead{1L1S} & \colhead{1L2S} & \colhead{2L1S(close)} & \colhead{2L1S(wide)}}
% \startdata
% $\chi^2_{raw}$ & 11143.41 & 11069.46 & 10687.49 & 10709.22  \\
% $\ln \mathcal{Z}_{raw}$ & -5573.60 & -5556.46 & -5378.75 & -5352.14 \\
% \hline
% $\chi^2_{renorm}$ & 1301.13 & 1301.28 & 1226.30 & 1226.08 \\
% $\ln \mathcal{Z}_{renorm}$ & -661.76 & -665.16 &  -644.78 & -644.64 \\
% \hline
% $\Delta \chi^2$ & 0.00 & 0.15 & 74.83 & 75.05 \\
% $\Delta \ln \mathcal{Z}$ & 0.0 & -3.40 & 16.98 & 17.12 \\
% \enddata
% \end{deluxetable}

\section{PROPERTIES OF THE LENSING SYSTEM\label{propr}}

In the previous sections we presented the results from the light-curve fitting, from which
our sources of physical information are $t_{\rm E}$ and $\rho_\ast$. Also, we could not break the
outer/inner geometry, but fortunately the fitting showed that the majority 
of the parameters are similar within their uncertainties (see Table~\ref{tab:param}), 
in particular the crossing time $t_{\rm E}$. Doing a direct comparison of the posterior 
distributions for both solutions,  it is clear that they also share a similar distribution
of $\rho_\ast$ as seen { in Figure~\ref{fig:te}, despite} the $s$ distributions being different.
This becomes important because we can proceed to the Galactic model expecting a 
priori that both solutions will share similar characteristics. 

\subsection{Bayesian Analysis}

We performed a Bayesian analysis to characterize the physical properties of the system.
As remarked above, our source of information from the light curve comes only 
from the distributions of $t_{\rm E}$ and $\rho_\ast$, which
restricts how much information we can gather. In other words, we need to limit the number
of parameters used in the analysis to the most basic. The main parameters are the total mass of the 
lens, $M_{L}$, the distance to the lens, $D_L$, the proper motion, $\mu_{rel}$, 
and the source distance, $D_S$. Nevertheless, during the formulation of the analysis,
the effective transverse velocity $v = D_L\mu_{rel} $ will be used as an
intermediate step during the parameter sampling.

Our analysis resembles a hierarchical Bayes~\citep[see e.g.,][]{gelman2003bayesian},
\begin{equation}
    p(\Theta|y) \propto p(\Theta)\int{p(t_{\rm E},\rho_\ast|\Theta)p(y|t_{\rm E},\rho_\ast)}dt_{\rm E} d\rho_\ast\, ,
\end{equation}\label{eq:propr.1}
where
\begin{eqnarray}
{\Theta} & \equiv & (M_L,D_L,D_S,v)  \\
y & \equiv & \text{data} \, .
\end{eqnarray}\label{eq:propr.2}

{This describe the probability 
distribution function for $t_{\rm E}$ and the 
weights based on the distribution of $\rho_\ast$, which is shown in Figure~\ref{fig:te}.}
{The prior sampling distribution for the physical quantities is $p(\Theta)$ , which} includes the information of the Galactic model.
For the analysis, $t_{\rm E}$ and $\rho_\ast$ {act as constraints,} and they are hidden
in $p(t_{\rm E},\rho_\ast|\Theta)$, which is the probability of a value of $t_{\rm E}$ and $\rho_\ast$ given by the physical parameters.

For this purpose we took a combined { approach in which we perform}
an MCMC simulation to obtain the posterior distribution from the
Galactic modelling. Ideally, we would describe the 
process in observable variables similar to~\cite{Batista2011MOA-2009-BLG-387Lb:Dwarf,Yee2012MOA-2011-BLG-293Lb:DETECTIONS}
or~\cite{Jung2018OGLE-2017-BLG-1522:Bulge}, but in this case
we are only able to relate two observables to the physical parameters, $t_{\rm E}$ and $\rho_\ast$.
We therefore take advantage of the Monte Carlo simulation to take care of the
marginalization. We 
used two different sampling algorithms in order to verify that our results contain no algorithmic bias. Independently, 
we ran the Emcee sampler~\citep{Foreman-Mackey2013EmceeHammer} which
uses the standard Metropolis-Hasting sampling method, and 
Dynesty~\citep{Higson2018DynamicCalculation} which uses Nested Sampling. 
After convergence, both algorithms arrived at the same posterior distributions.  

Our approach is similar to that of~\cite{Yoo2004ConstraintsOGLE2003BLG423}, but we describe our Galactic model
and the fact that we can define the probability of 
a physical property given an observed parameter, { e.g.,} $p(\Theta|t_{\rm E}^{(0)})$,
as presented in~\cite{Dominik1998EstimatingEvent,Albrow2000Detection97BLG41}. 
% Considering that, a-priori, we assumed equal probability for each value of $t_{\rm E}$ and $\rho_\ast$, for
% the MCMC sampling process  the  probabilities $p(\theta|t_{\rm E},\rho_\ast) \leftrightarrow p(t_{\rm E},\rho_\ast|\theta)$ become interchangeable.

For the Galactic model, the lens mass function, $\Phi_{\log m}$, is
a power law or Gaussian distribution depending  on the mass ranges according to~\cite{Chabrier2003GalacticFunction}. The mass density distribution, $\Phi_x(x)$,
considers the disk, bulge or both depending on the case, where the disk is modelled by
a double exponential with $0.3$ kpc  and $1.0$ kpc as the scale heights of a thin and thick disk perpendicular to the Galactic plane, and the corresponding column mass
densities are $\Sigma_{\text{thin}}= 25 \, \text{M}_\odot\text{pc}^{-2}$ and 
$\Sigma_{\text{thick}}= 35 \, \text{M}_\odot\text{pc}^{-2}$. For the bulge we adopt
a model of a { barred bulge that } is tilted by an angle of $20^{\circ}$~\citep[see][]{Grenacher1999,Han1995TheMeasurements}. The probability
density of the absolute effective velocity, $\Phi_\nu(\nu,x)$, assumes Gaussian 
distributions, for which isotropic velocity dispersions are {assumed for the Galactic
disk} and bulge, and the values of $\sigma^{\text{{ disk}}}=30 \, \text{km}\text{s}^{-1}$ and
$\sigma^{\text{bulge}}=100 \, \text{km}\text{s}^{-1}$are adopted. While the velocity mean for Bulge objects is assumed
purely random, the { disk} lenses rotation velocity can described by a NFW model~\citep[see][]{NFW1997}. The precise equations for each 
component of the Galactic model can be found in the appendices of~\cite{Dominik2006StochasticEvents}.

Therefore, the probability of our assumed Galactic model~(galactic prior) for the lens is in the form
% \begin{equation}\label{eq:propr.3}
%     p(t_{\rm E},\rho_\ast|\theta) \propto (p_{bulge\,lens} \cup p_{disk\,lens})\cap p_{source}\, , 
% \end{equation}
% where the lens distributions are constructed as
\begin{eqnarray}\label{eq:propr.4}
p_i(m,\zeta,x) \propto \Phi_{\log m}(\log m)\Phi_\zeta(\zeta,x)\Phi_x(x) \, , 
\end{eqnarray}
{selecting from bulge or disk populations accordingly.} The parameters in this prescription are the mass 
$m = M_L/\text{M}_\odot$, the fractional lens-source distance, $x = D_L/D_S$, and the effective transverse velocity 
$\zeta = v/v_c$, where $v_c = 100 \, \text{km}\,\text{s}^{-1}$ is a scaling constant to 
keep the velocity dimensionless.
These distributions and prescriptions can also be found in~\cite{Dominik2006StochasticEvents},
and, as mentioned above, the assumed properties are described in its appendix. 

The source distance probability distribution is defined as
\begin{equation}\label{eq:propr.6}
p_{source}(D_S) = \frac{D_S^{\gamma}\rho(D_S)}{\int^{D_{s,max}}_0{D_S^{\gamma}\rho(D_S)}dD_S},
\end{equation}
where $\rho(D_S)$ is the density of objects at the source distance as defined 
in~\cite{Dominik2006StochasticEvents}. We adopted this distribution as we { do not} 
have any information on the source location, and
we based its definition on the~\cite{Zhu_2017} argument, but we use a { value} of $\gamma = 1$.
For our calculations, we assumed that the source was part of the bulge population.

Therefore, the Galactic {prior, which has the Galactic model information, is} sampled from the
probability distribution
\begin{equation}\label{eq:galprior}
    p(\Theta) \propto  \Omega(m,\zeta,x) p_i(m,\zeta,x) p_{source}(D_S).
\end{equation}
The lens for bulge or disk populations are selected accordingly. Meanwhile, the
information of the prior is weighted by
\begin{equation} \label{eq:galprior2}
    \Omega(m,\zeta,x) \propto (M/M_\odot)^{1/2}\zeta \sqrt{x(1-x)}.
\end{equation}
\citep{Dominik2006StochasticEvents}.
The parameter $t_{\rm E}$ then becomes 
intrinsic and is defined as
\begin{eqnarray}\label{eq:propr.5}
t_{\rm E}(m,\zeta,x,D_S) &=& \frac{2r_{E,\odot}\sqrt{m x(1-x)}}{\zeta{v_c}}\, ,
\end{eqnarray}
where
\begin{eqnarray}\label{eq:propr.5b}
r_{E,\odot} &=& \sqrt{\frac{G\text{M}_\odot}{c^2}D_S}
\end{eqnarray}
is a scale length defined as the Einstein radius of a solar mass lens located
half-way between the observer and source.

The angular source size parameter,
\begin{equation}\label{eq:rhoast}
    \rho_\ast(m,x,D_S) = \theta_\ast/\theta_{\rm E}(m,x,D_S)\, , 
\end{equation}
where $\theta_\ast$ is the source angular radius given in the previous section and $\theta_{\rm E}(m,x,D_S)$ is the Einstein angle for a MCMC realization. 

Given that $\rho_\ast$ and $t_{\rm E}$ are derived from the prior parameters,
their joint probability
\begin{equation}\label{eq:probtheta} 
     p(t_{\rm E},\rho_\ast|\Theta) \propto \delta(t_{\rm E}-t_{\rm E}(\Theta)) \delta(\rho_\ast - \rho_\ast(\Theta))\, .
\end{equation}

Additionally, there is a weight given by the information contained in
the probability distribution $p(y|t_{\rm E},\rho_\ast)$, which corresponds to the posterior distributions of our solution from our light-curve fitting. 

The simplest to represent is the value of $t_{\rm E}$, as it is well constrained and reduces simply to
\begin{equation}\label{eq:probte}
     p(y|t_{\rm E}) \propto \exp\left(-\frac{(t_{\rm E}-t_{E,{best}})^2}{2\sigma_{t_{\rm E}}^2}\right)\, .
\end{equation}
Here $t_{E,{\rm best}}$ and $\sigma_{t_{\rm E}}$ correspond to the values from the best fit of the distribution from the light-curve fitting.

\begin{figure}
 \includegraphics[width = 0.99\columnwidth]{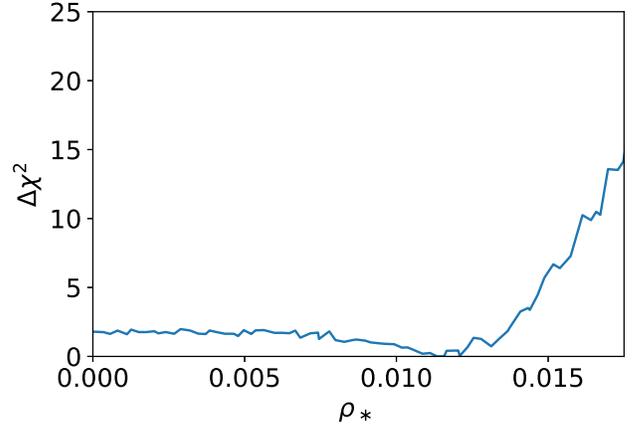}
%\plotone{galmod}
\caption{ Minimum  $\Delta \chi ^2$ as a function of $\rho_\ast$ using the chains derived from the light-curve fitting MCMC. This shows that $\rho_\ast$ is weakly detected as 
there is clearly an upper limit for it, but the difference from a flat region on the smaller values is only $\Delta \chi^2\approx 1.7$. 
\label{fig:rhofunc}}
\end{figure}

For the case of $\rho_\ast$, it is not as simple because the constraint is not well approximated as a Gaussian. We introduce the information by the 
use of $\Delta \chi^2 = \chi^2 -\chi^2_{min}$ of each of the light-curve fitting samples. From the lower envelope of these
samples, binned in $\rho_\ast$, we obtain
a numerical function $\Delta \chi^2(\rho_\ast)$, see Figure~\ref{fig:rhofunc}. 
The probability of $\rho_\ast$ is then given by

\begin{equation}\label{eq:probrho}
     p(y|\rho_\ast) \propto \exp\left(-\frac{\Delta \chi^2(\rho_\ast)}{2}\right)\, .
\end{equation}
The previous weights are combined as 
\begin{equation}\label{eq:weights}
     p(y|t_{\rm E},\rho_\ast) \propto p(y|\rho_\ast) p(y|t_{\rm E})\,.
\end{equation}
The information of the Galactic model is introduced into Equation~(\ref{eq:weights}) by the restriction imposed by the Dirac $\delta$ function in Equation~(\ref{eq:probtheta}).

The {combination of the prior and these weights gives} the desired posterior probability for the physical parameters, where $\mu_{rel}$ is appropriately obtained from $\zeta$. We note 
that both outer and inner models lead to the same final results. This is expected given the similarity of the $t_{\rm E}$ and $\rho_\ast$ distributions.
% , and secondly as a superposition of Gaussian functions as described in~\cite{Shin2019Two2017}. The fitted parameter values for the superposition method are given in Table~\ref{tab:fit}.

% \begin{deluxetable}{c|c|c|c|c|c|c}
% \tablecaption{Double Gaussian fitting parameters.\label{tab:fit}}
% \tablehead{ \colhead{Solution} & \colhead{$a_1$} & \colhead{$a_2$}
% & \colhead{$\mu_1$}&\colhead{$\mu_2$}&\colhead{$\sigma_1$}&\colhead{$\sigma_2$}}
% \startdata
% Wide & 3.464 &  0.536 &  4.867 &  4.820 & 0.111 & 0.024 \\
% Close & 0.293 & 3.312 & 4.736 & 4.889 & 0.0585 & 0.115
% \enddata
% \end{deluxetable}

\begin{figure}
 \includegraphics[width = 0.99\columnwidth]{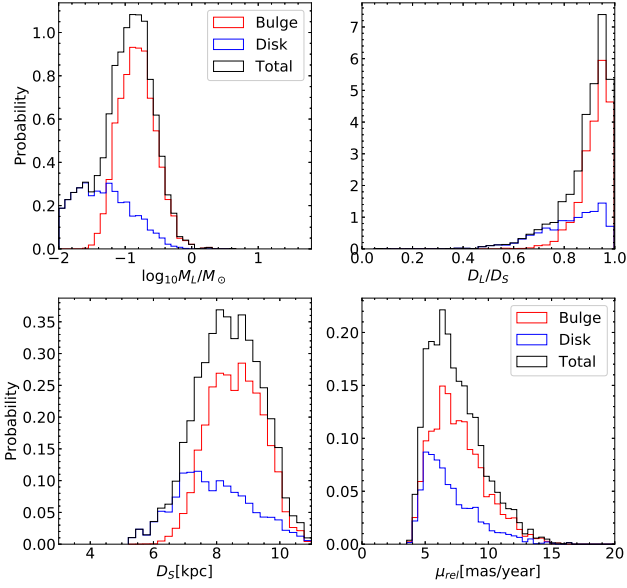}
%\plotone{galmod}
\caption{ Posterior probability distributions from the Bayesian analysis for the 
lens mass $log_{10}M_L/M_\odot$, the ratio of the lens and source distance $D_L/D_S$, the source distance $D_S$ and $\mu_{rel}$. In addition to the total, the four panels show also the separate bulge and disk distributions, normalized to the total. 
\label{fig:galmod}}
\end{figure}

\begin{figure}
 \includegraphics[width = 0.99\columnwidth]{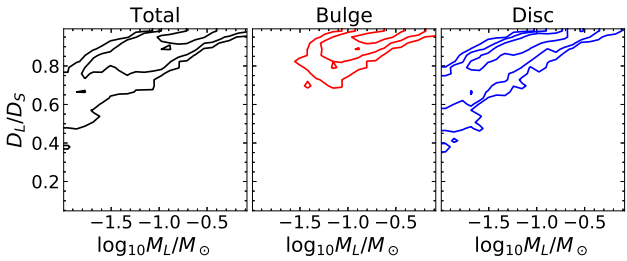}
\caption{1, 2, and 3-sigma contour levels for the distributions of the 
lens mass $log_{10}M_L/M_\odot$ and the ratio of the lens and source distance $D_L/D_S$ . In addition to the total, the panel show also the separate bulge and disk distributions. 
\label{fig:Mds}}
\end{figure}
% Bayesian analysis posterior results showing the resulting normalized distributions from the analysis. The blue, cyan and black contours represent the 1, 2 and 3$\sigma$ 
% contours in two dimensions respectively. The green and red curves represent the priors for Bulge and the black the total as evaluated at the specified value. These are used as a guide only given that, except for the mass, the prior distributions are functions of several parameters. 
\begin{deluxetable}{c c c c}
\tablecaption{Estimators from the Bayesian results \label{tab:galmod}}
\tablehead{ \colhead{$\log_{10}{M_L/M_\odot}$} & \colhead{$D_L/D_S$} & \colhead{$D_S$[kpc]}
& \colhead{$\mu_{rel}$[mas/year]}}
\startdata
${-0.94}^{+0.34}_{-0.47}$ & ${0.92}^{+0.05}_{-0.12}$ & ${8.37}^{+1.06}_{-1.07}$ &  ${7.02}^{+2.48}_{-1.68}$
\enddata
\end{deluxetable}

%old
% \begin{deluxetable}{c c c c}
% \tablecaption{Estimators from the Bayesian results \label{tab:galmod}}
% \tablehead{ \colhead{$\log_{10}{M_l/M_\odot}$} & \colhead{$D_l/D_S$} & \colhead{$D_S$[kpc]}
% & \colhead{$v/v_c$}}
% \startdata
% ${-0.91}_{-0.63}^{+0.64}$ & ${0.79}_{-0.56}^{+0.18}$ & ${8.38}_{-1.58}^{+1.97}$ &  ${3.47}_{-1.83}^{+5.77}$
% \enddata
% \end{deluxetable}

\subsection{Resulting properties\label{properties}}

The distributions obtained from the Galactic modelling are presented in
{ Figure~\ref{fig:galmod},} and the corresponding estimators for the median and error
values are in Table~\ref{tab:galmod}. 
As noted above, the outer and inner solutions produce 
indistinguishable distributions.

\begin{figure}[H]
    \includegraphics[width = 0.99\columnwidth]{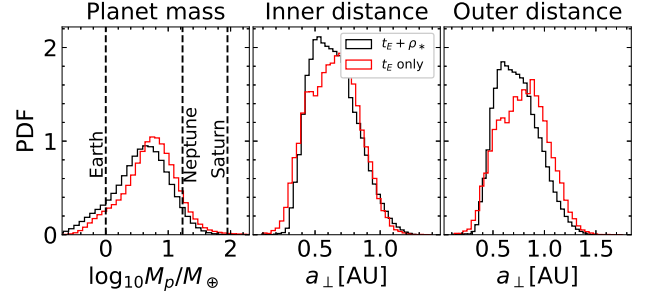}
    %\plotone{planet}
    \caption{Derived distributions for the planet mass and projected separation, $a_\perp$[AU], in Astronomical Units for outer and inner solutions. The red line shows the change
    caused by removing information about $\rho_\ast$ from the Galactic model evaluation.
    }
    \label{fig:planet}
\end{figure}

% \begin{figure}[H]
%     \includegraphics[width = 0.99\columnwidth]{dist}
%     %\plotone{planet}
%     \caption{Derived distributions for projected
%     separation, d[AU], in Astronomical Units for Wide and Close solutions. 
%     }
%     \label{fig:dist}
% \end{figure}

From the resulting distributions of { the basic four parameters, the lens mass, lens distance, source distance and proper motion,} we can propagate the
information to derive the physical mass of the binary lens components and their physical separation. Additionally, the host-mass {distribution is essentially identical to} the total lens-mass distribution 
in Figure~\ref{fig:galmod}.
In this case, the binary-lens mass ratio is small, and the host consists of a low mass
star or brown dwarf with a super-Earth companion.
As mentioned previously, given the overlapping of the distributions of $t_{\rm E}$ and $q$ 
for the inner and outer solutions, the Bayesian analysis returns indistinguishable 
distributions, and for these reasons it is justifiable to say that both give
the same derived planet mass distribution. This is 
not the case for the projected separation between host and planet, which has a
small but not negligible difference.
The derived distributions are shown in Figure~\ref{fig:planet} and their estimators
in Table~\ref{tab:planet}. 

Furthermore, as mentioned at the end of~\ref{subsec:model}, the value for $\rho_\ast$ is weakly constrained.  The relatively broad and highly non-Gaussian form of this constraint is shown in  Figure~\ref{fig:rhofunc}. For comparison purposes, in Figure~~\ref{fig:planet}, we show alongside the complete Galactic modeling, which includes the information 
of $t_{\rm E}$ and $\rho_\ast$, the derived distributions produced by removing the information about $\rho_\ast$ and using only $t_{\rm E}$.
In such a case, the planet mass distribution is shifted a little higher, and the projected separation distributions are a little broader.
Their estimators are also included in Table~\ref{tab:planet}. 

\begin{deluxetable}{c c c c }
\tablecaption{Estimators for the planet mass and projected separation \label{tab:planet}}
\tablehead{ & \colhead{$\log_{10}{M_p/M_\oplus}$} & \colhead{$a_\perp$(Inner)} & \colhead{$a_\perp$(Outer)}}
\startdata
 $t_{\rm E} + \rho_\ast$ & ${0.60}^{+0.40}_{-0.48}$ & ${0.63}^{+0.20}_{-0.17}$ AU & ${0.72}^{+0.23}_{-0.19}$ AU \\
$t_{\rm E}$ only & ${0.73}^{+0.37}_{-0.45}$ & ${0.64}^{+0.19}_{-0.21}$ AU & ${0.80}^{+0.23}_{-0.26}$ AU
\enddata
\end{deluxetable}

%old
% \begin{deluxetable}{c c c }
% \tablecaption{Estimators for the planet mass and projected separation \label{tab:planet}}
% \tablehead{ \colhead{$\log_{10}{M_p/M_\oplus}$} & \colhead{$d$(Close)} & \colhead{$d$(Wide)}}
% \startdata
% ${0.80}^{+0.38}_{-0.35}$ & ${0.86}^{+0.74}_{-0.27}$ AU & ${1.10}^{+1.01}_{-0.35}$ AU
% \enddata
% \end{deluxetable}

% The close and wide solutions for the planet in BLG0816 both imply that the secondary lens is 
% a super-Earth/Sub-Neptune planet with a mass $M_{\mathrm{planet}} = {6.34}^{+8.96}_{-3.54}\mathrm{M_\oplus}$, while
% the host is a dwarf star or brown-dwarf with mass
% $ M_{\mathrm{host}} = {0.13}^{+0.18}_{-0.07}\mathrm{M_\odot}$.
% The projected separation between the star and planet is ${0.86}^{+0.74}_{-0.27}$ AU and
% ${1.10}^{+1.01}_{-0.35}$ AU for the close and wide solutions respectively. The lens 
% system is most likely to be at a distance of ${6.60}^{+1.45}_{-3.13}$ kpc,  which is relatively unconstrained but most-probably in the Bulge.

The inner and outer solutions for the planet in { OGLE-2018-BLG-0667} both imply that the secondary lens is 
a super-Earth/Sub-Neptune planet with a mass $M_{\mathrm{planet}} = {3.96}^{+5.88}_{-2.66}\mathrm{M_\oplus}$, while
the host is a dwarf star or brown-dwarf with mass
$ M_{\mathrm{host}} = {0.12}^{+0.14}_{-0.08} M_\sun$.

The projected separation between the star and planet is ${0.63}^{+0.20}_{-0.17}$ AU and
${0.72}^{+0.23}_{-0.19}$ AU for the inner and outer { solutions, respectively.} The lens 
system is estimated to be at a distance of ${7.58}^{+1.15}_{-1.35}$ kpc, {with a 66.9\% probability of lying in the bulge (Figure~\ref{fig:galmod}).}

{
For completeness, we note the physical implications of the solutions corresponding to the two local minima discussed in Section~\ref{subsec:local}. Since these are degenerate with each other in $q$, $t_{\rm E}$ and $\rho_*$, they imply the same primary and secondary lens mass and distance. The local minima would imply a total lens mass of 0.1 $M_\odot$ and a planet mass of 6.43 $M_\oplus$, values very similar to the favoured solutions, but a lens located at a distance of 0.99 $D_S$. The projected separation of the planet from its host would be smaller, either 0.20 or 0.22 AU, for the inner and outer local minima respectively. 
 }

\section{DISCUSSION}

OGLE-2018-BLG-0677Lb has the lowest $\Delta \chi^2$ relative to the best 1L1S model of any planet
securely detected by microlensing. One of the first microlensing planets, OGLE-2005-BLG-0390 
\citep{Beaulieu2006} {also had one of the lowest improvements relative to 1L1S, $\Delta\chi^2=960$.  One reason such seemingly high formal thresholds are generally required for secure microlensing planet detections is that ``bumps'', particularly those without clear caustic features, can in principle be produced by 1L2S models~\citep{Gaudi1998DistinguishingPerturbations,Jung_2017, shin2019dege}.  

In the case, of OGLE-2005-BLG-390, the 1L2S model was ruled out by just $\Delta\chi^2=46$.
Moreover, \citet{ob171434} examined whether OGLE-2005-BLG-390 would have been detectable if the planet-host mass ratio were lower.  They concluded that at factors $q^\prime/q<0.63$ it would not have been because the resulting $\Delta\chi^2=\chi^2({\rm 1L2S})-\chi^2({\rm 2L1S})\leq 13$ would have been too marginal to claim reliable detection of a planet.
Note that this threshold for unambiguous identification corresponds to
$\Delta\chi^2=\chi^2({\rm 1L1S})-\chi^2({\rm 2L1S})\simeq 457$, which is almost ten times the value for OGLE-2018-BLG-0677.} 

{However,  OGLE-2018-BLG-0677 is detected primarily through a
dip }in the light curve, which cannot be reproduced by any 1L2S scenario, or any other higher-order microlensing effect. The only other possible cause for the observed dip is a systematic error in the photometry. This is extremely unlikely because the signal is detected in the data from two overlapping KMTC fields that were reduced independently, and is confirmed by two data points from the OGLE telescope, located at a different observatory, and with images reduced by different software.

With a relative lens-source proper motion greater than 5 mas/yr, the separation between lens and source will be sufficient for them to be separately resolved within a decade with the advent of IR adaptive optics imaging on {either present-day or under-construction extremely large telescopes.} We have no detection of flux from the blend in the current {event, and so we expect} that the lens is at least ten times fainter than the source, $I_{\rm lens} >  22$. {This implies that to be confident that a non-detection of the lens implies a non-luminous (i.e., brown-dwarf) lens, these observations should be carried out after the lens and source have separated by at least 1.5 FWHM, i.e., $12\,{\rm yr} (\lambda/1.6\,\mu{\rm m})/(D/10\,{\rm m})$ after $t_0$ (i.e., 2018).  Here $\lambda$ is the wavelength of the observations and $D$ is the diameter of the mirror.}

\section{SUMMARY} \label{sec:discuss}
We have presented an analysis of the microlensing event OGLE-2018-BLG-0677/OGLE-2018-BLG-0680/KMT-2018-BLG-0816. 

The light curve of the event exhibits a small dip, soon after peak on an otherwise-smooth { Paczy\'nski-like} 
magnification profile. {The dip lasts for $\sim 3.5$ hours and is followed by a smaller bump lasting $\sim 1.5$ hours.  These features are well traced by the KMTNet CTIO observations from two telescope pointings and are confirmed by the OGLE-2018-BLG-0680 light curve.}

We have fitted several models to the light curve and obtained their $\chi^2$ and evidence. The light curves for
1L2S and 1L1S models { are virtually identical} and do not  reproduce the anomalous feature in the observations. 

%%We normalized  with the different solutions, and 
%%decided to use the binary lens close solution as it had the lowest $\chi^2$ and the best-fit results remained %%unchanged despite the change in normalization. Although, the difference in
%%the minimum was considerable initially, it was reduced to $\Delta\chi^2\approx 70$ after the renormalization which %%still prefers the binary over the single lenses solutions, but it is
%%a small gap which would need the use of an extra data set, e.g. OGLE,  to confirm a more robust
%%decision to exclude any possible systematic error on the dataset which would affect the preference for the binary %%solutions.

We  also presented the fit for binary lens solutions.
These suffer from a two-fold inner-outer(close-wide) degeneracy, but we found that
the ratio of masses was similar for both, indicating that the
lens is composed of a star plus planet system.
Formally, the planet is detected with a $\Delta \chi^2 \approx 46$ and $\Delta \ln \mathcal{Z} = 5$
relative to the single lens models.

Using the binary  solutions we performed a Bayesian analysis and found
that outer and inner solutions agree with the same distribution
for the lens mass, lens distance { and proper motion.}

\begin{figure}[H]
    \includegraphics[width = 0.99\columnwidth]{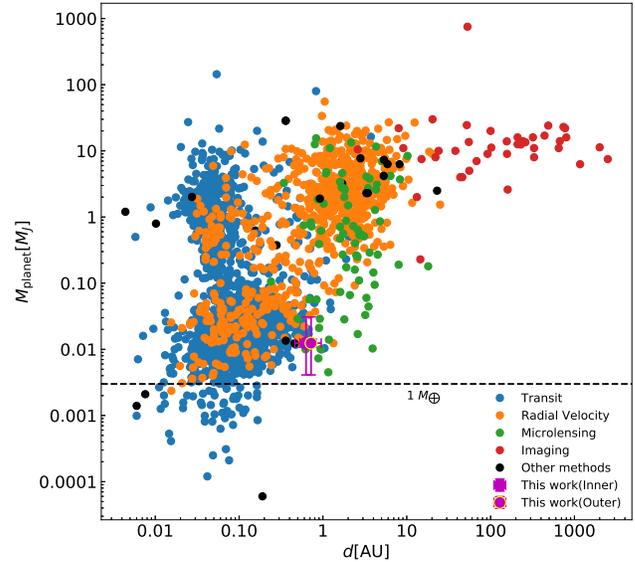}
    %\plotone{methods}
    \caption{Masses and orbital radii of confirmed exoplanets detected via different methods. 
    A dashed line indicates where Earth-mass planets would lie. The data
    for the confirmed exoplanets can be found in NASA Exoplanet Archive (https://exoplanetarchive.ipac.caltech.edu).}
    \label{fig:methods}
\end{figure}

The derived planet mass is
$M_{\mathrm{planet}} = {3.96}^{+5.88}_{-2.66}\mathrm{M_\oplus}$, one of the lowest yet detected 
by microlensing.
In  Figure~\ref{fig:methods} we use data from the \href{https://exoplanetarchive.ipac.caltech.edu}{NASA Exoplanet Archive}
to show 
the mass and separation of confirmed exoplanets found with different methods such as microlensing, radial velocity or transits. 
(An example of how the mass is derived for the latter can be found { in \citealt{Espinoza2016DISCOVERYMISSION}.) }

It can be seen that the planet mass and
distance reported in this work are consistent with the lower limit
of masses that have been found previously through microlensing, and also
consistent with their separation range. This event
reinforces the importance of high-cadence microlensing observations,
which allow us to detect and characterize anomalies due to low-mass planets.

\acknowledgments
{
AHM and MDA are supported by the Marsden Fund under contract UOC1602.

The OGLE project has received funding from the National Science Centre, Poland, grant MAESTRO 2014/14/A/ST9/00121 to AU.

W.Z. acknowledges support by the National Science Foundation of China (Grant No. 11821303 and 11761131004) % Work by AG was supported by AST-1516842 from the US NSF and by JPL grant 1500811.
AG received support from the European  Research  Council  under  the European  Union's Seventh Framework Programme (FP 7) ERC Grant Agreement n. [321035] 
This research has made use of the KMTNet system operated by the Korea Astronomy and Space Science Institute (KASI) and the data were obtained at three host sites of CTIO in Chile, SAAO in South Africa, and SSO in Australia.
Work by C.H. was supported by the grants (2017R1A4A1015178 and 2019R1A2C2085965)
of National Research Foundation of Korea.

}

%% Bibliography section

\bibliographystyle{aasjournal}
\bibliography{references}

\end{document}